\documentclass[twocolumn,showpacs,superscriptaddress,pre]{revtex4}

\usepackage{graphicx}
\usepackage{dcolumn}
\usepackage{bm}
\usepackage{amsmath}
\usepackage{amssymb}
\usepackage{times}
\usepackage{color}

\begin{document}

\title{Peer pressure: enhancement of cooperation through mutual punishment}

\author{Han-Xin Yang}\email{hxyang01@gmail.com}
\affiliation{Department of Physics, Fuzhou University, Fuzhou
350108, China}

\author{Zhi-Xi Wu}
\affiliation{Institute of Computational Physics and Complex
Systems, Lanzhou University, Lanzhou, Gansu 730000, China}

\author{Zhihai Rong}
\affiliation{Web Sciences Center, University of Electronic
Science and Technology of China, Chengdu 610054, China}

\author{Ying-Cheng Lai}
\affiliation{School of Electrical, Computer and Energy Engineering,
Arizona State University, AZ 85287, USA}

\begin{abstract}

An open problem in evolutionary game dynamics is to understand the
effect of peer pressure on cooperation in a quantitative manner.
Peer pressure can be modeled by punishment, which has been proved to
be an effective mechanism to sustain cooperation among selfish
individuals. We investigate a symmetric punishment strategy, in
which an individual will punish each neighbor if their strategies
are different, and vice versa. Because of the symmetry in imposing
the punishment, one might expect intuitively the strategy to have
little effect on cooperation. Utilizing the prisoner's dilemma game
as a prototypical model of interactions at the individual level, we
find, through simulation and theoretical analysis, that proper
punishment, when even symmetrically imposed on individuals, can
enhance cooperation. Besides, we find that the initial density of
cooperators plays an important role in the evolution of cooperation
driven by mutual punishment.

\end{abstract}

\date{\today}

\pacs{02.50.Le, 87.23.Kg, 87.23.Ge}

\maketitle

\section{Introduction} \label{sec:intro}

Cooperation is ubiquitous in biological, social and economical
systems~\cite{1}. Understanding and searching for mechanisms
that can generate and sustain cooperation among selfish individuals
remains to be an interesting problem. Evolutionary game theory
represents a powerful mathematical framework to address this
problem~\cite{2,3}. Previous
theoretical~\cite{the0,the1,the2,the3,the4,the5,the6,the7}
and experimental~\cite{exp1,exp2,exp3,exp4,exp5,exp6,exp7,exp8}
studies showed that, for evolutionary game dynamics in spatially
extended systems, punishment is an effective approach to
enforcing the cooperative behavior, where the punishment can
be imposed on either cooperators or defectors. The agents that
get punished bear a fine while the punisher pays for the cost
of imposing the punishment~\cite{cost1,cost2}. In  existing
studies, individuals who hold a specific strategy (usually
defection) are punished.

In realistic situations, punishment can be mutual and the
strategy would typically depends on the surrounding environment,
e.g., on neighbors' strategies. An example is ``peer pressure.''
Previous psychological experiments demonstrated that, an individual
tends to conglomerate (fit in) with others in terms of behaviors or
opinions~\cite{behavior}. Dissent often leads to punishment
either psychologically or financially, or both, as human individuals
attempt to attain social conformity modulated by peer
pressure~\cite{behavior,dissent1,dissent2}. To understand
{\em quantitatively} the effect of peer pressure on cooperation
through developing and analyzing an evolutionary game model is the
main goal of this paper. In particular, we propose a mechanism
of punishment in which an individual will punish neighbors who hold
the opposite strategy, regardless of whether they are cooperators
or defectors.

Differing from previous models where additional strategies of
punishment were introduced, in our model there are only two
strategies (pure cooperators and pure defectors). More importantly,
the punishment is mutual in our model, i.e., individual $i$ who
punishes individual $j$ is also punished by $j$, so the cost of
punishment can be absorbed into the punishment fine. Because of this
symmetry at the individual or ``microscopic'' level, intuitively one
may expect the punishment not to have any effect on cooperation.
Surprisingly, we find that symmetric punishment can lead to
enhancement of cooperation. We provide computational and heuristic
arguments to establish this finding.

\section{Model} \label{sec:model}

Without loss of generality, we use and modify the classic prisoner's
dilemma game (PDG)~\cite{pdg} to construct a model to gain
quantitative understanding of the effect of peer pressure on
cooperation by incorporating our symmetric punishment mechanism. In
the original PDG, two players simultaneously decide whether to
cooperate or defect. They both receive payoff $R$ upon mutual
cooperation and payoff $P$ upon mutual defection. If one cooperates
but the other defects, the defector gets payoff $T$ while the
cooperator gains payoff $S$. The payoff rank for the PDG is $T > R >
P > S$. As a result, in a single round of PDG, mutual defection is
the best strategy for both players, generating the well-known social
dilemma. There are different settings of payoff
parameters~\cite{payoff1,payoff2}. For computational
convenience~\cite{nowak}, the parameters are often rescaled as
$T=b>1$, $R=1$, and $P=S=0$, where $b$ denotes the temptation to
defect.

In their pioneering work, Nowak and May included spatial structure
into the PDG~\cite{nowak}, in which individuals play games only with
their immediate neighbors. In the spatial PDG, cooperators can
survive by forming clusters in which mutual cooperation outweigh the
loss against defectors~\cite{cluster1,cluster2,cluster3,cluster4}.
In the past decade, the PDG has been extensively studied for
populations on various types of network
configurations~\cite{spatial1,spatial2,spatial3}, including regular
lattices~\cite{lattice1,lattice2,lattice3,lattice5}, small-world
networks~\cite{small1,small2}, scale-free
networks~\cite{scale1,scale2,scale3,scale4}, dynamic
networks~\cite{dynamics0,dynamics1,dynamics2,dynamics3}, and
interdependent networks~\cite{interdependent1}.

Our model is constructed, as follows. Player $x$ can take one of two
strategies: cooperation or defection, which are described by
\begin{equation}\label{1}
s_{x} =\left(
     \begin{array}{c}
       1 \\
       0 \\
     \end{array}
   \right)\mathrm{or}\left(
     \begin{array}{c}
       0 \\
       1 \\
     \end{array}
   \right),
\end{equation}
respectively. At each time step, each individual plays the PDG with
its neighbors. An individual will punish the neighbors that hold
different strategies. The accumulated payoff of player $x$ can thus
be expressed as
\begin{equation} \label{2}
P_{x}=\sum_{y\in \Omega_{x}}[s_{x}^{T}Ms_{y}-\alpha
(1-s_{x}^{T}s_{y})],
\end{equation}
where the sum runs over the nearest neighbor set $\Omega_{x}$ of
player $x$, $\alpha$ is the punishment fine, and $M$ is the rescaled
payoff matrix given by
\begin{equation}\label{3}
M=\left(
    \begin{array}{cc}
      1 & 0 \\
      b & 0 \\
    \end{array}
  \right).
\end{equation}

Initially, the cooperation and the defection strategies are randomly
assigned to all individuals in terms of some probabilities: the
initial densities of cooperators and defectors are set to be
$\rho_0$ and $1-\rho_0$ respectively. The update of strategies is
based on the replicator equation~\cite{replication} for well-mixed
populations and the Fermi rule~\cite{update0} for structured
populations.

\section{Results for well-mixed populations} \label{sec:well-mixed}

In the case of well-mixed populations, i.e., a population with no
structure, where each individual plays with every other, the
evolutionary dynamics is determined by the replication equation of
the fraction of the cooperators $\rho$ in the
population~\cite{replication}:
\begin{equation}\label{3.1}
\frac{d \rho}{dt}=\rho(1-\rho)(P_{c}-P_{d}),
\end{equation}
where $P_{c}=\rho-(1-\rho)\alpha$ is the rescaled payoff of a
cooperator and $P_{d}=\rho b-\rho\alpha$ is the rescaled payoff of a
defector. The equilibria of $\rho$ can be obtained by setting $d
\rho/dt=0$. There exists a mixed equilibrium
\begin{equation}\label{3.2}
\rho_e = \frac{\alpha}{2\alpha+1-b},
\end{equation}
which is unstable. Provided that the initial density of cooperators
$\rho_{0}$ is different from 0 and 1, the asymptotic density of
cooperators $\rho_{c}=1$ if $\rho_{0}>\rho_{e}$, and $\rho_{c}=0$ if
$\rho_{0}<\rho_{e}$.

\begin{figure}
\begin{center}
\scalebox{0.4}[0.4]{\includegraphics{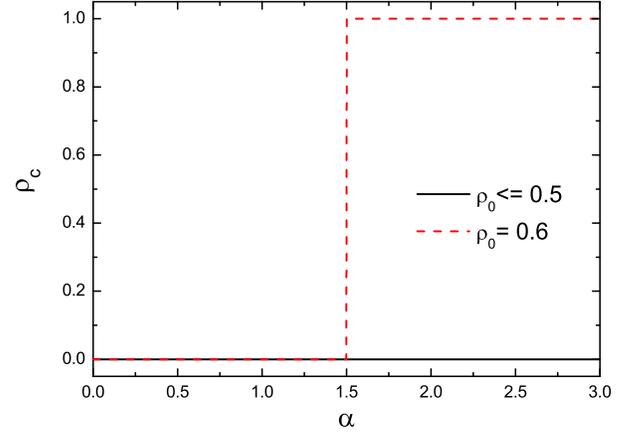}} \caption{(Color
online) Asymptotic density of cooperators $\rho_{c}$ as a function
of the punishment fine $\alpha$ for different values of the initial
density of cooperators $\rho_{0}$. The temptation to defect
$b=1.5$.} \label{fig0}
\end{center}
\end{figure}

\begin{figure}
\begin{center}
\scalebox{0.4}[0.4]{\includegraphics{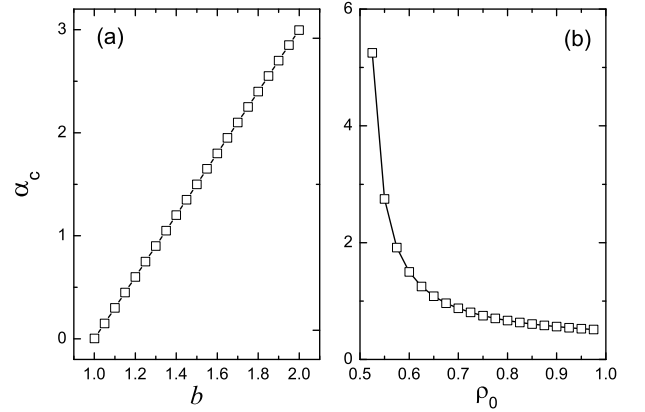}} \caption{(a) The
critical value of the punishment fine $\alpha_{c}$ as a function of
the temptation to defect $b$. The initial density of cooperators
$\rho_{0}=0.6$. (b) The dependence of $\alpha_{c}$ on $\rho_{0}$.
The temptation to defect $b=1.5$.} \label{fig0.1}
\end{center}
\end{figure}

Figure~\ref{fig0} shows the asymptotic density of cooperators
$\rho_{c}$ as a function of the punishment fine $\alpha$ for
different values of the initial density of cooperators $\rho_{0}$
when the temptation to defect $b=1.5$. From Eq. (\ref{3.2}), we note
that the mixed equilibrium $\rho_e$ definitely exceeds 0.5. As a
result, for $\rho_{0}\leq0.5$, $\rho_{c}$ is always zero regardless
of the values of the temptation to defect and the punishment fine.
However, for $0.5<\rho_{0}<1$, there exist a critical value of the
punishment fine (denoted by $\alpha_{c}$), below which cooperators
die out while above which defectors become extinct. According to Eq.
(\ref{3.2}), we obtain $\alpha_{c}$ as
\begin{equation}\label{3.3}
\alpha_{c} = \frac{(b-1)\rho_0}{2\rho_0-1}.
\end{equation}
For example, $\alpha_{c}=1.5$ when $\rho_{0}=0.6$ and $b=1.5$. From
Eq. (\ref{3.3}), one can find that $\alpha_{c}$ increases as the
temptation to defect $b$ increases but it decreases as the initial
density of cooperators $\rho_{0}$ increases, as shown in
Fig.~\ref{fig0.1}.

\section{Results for structured populations}

In a structured population, each individual plays the game only with
its immediate neighbors. Without loss of generality, we study the
evolution of cooperation on a square lattice, which is the simple
and widely used spatial structure. In the following, we use a
$100\times 100$ square lattice with periodic boundary conditions. We
find that the results are qualitatively unchanged for larger system
size, e.g., $200\times 200$ lattice.

In the following studies, we set the initial density of cooperators
$\rho_{0}=0.5$ without special mention. Players asynchronously
update their strategies in a random sequential
order~\cite{update0,update1,update2}. Firstly, player $x$ is
randomly selected who obtains the payoff $P_x$ according to
Eq.~(\ref{2}). Next, player $x$ chooses one of its nearest neighbors
at random, and the chosen neighbor $y$ also acquires its payoff
$P_y$ by the same rule. Finally, player $x$ adopts the neighbor's
strategy with the probability~\cite{update0}:
\begin{equation}\label{4}
W(s_{x}\leftarrow s_{y})=\frac{1}{1+\exp[-(P_y-P_x)/K]},
\end{equation}
where parameter $K$ characterizes noise or stochastic factors to
permit irrational choices. Following previous
studies~\cite{update0,update1,update2}, we set the noise level to be
$K=0.1$. (Different choices of $K$, e.g., $K=0.01$ and $K=1$, do not
affect the main results.)

The key quantity to characterize the cooperative behavior of the
system is the fraction of cooperators $\rho_{c}$ in some steady
state. All simulations are run for 30000 time steps to ensure that
the system reaches a steady state, and $\rho_{c}$ is obtained by
averaging over the last 2,000 time steps. Each time step consists of
on average one strategy-updating event for all players. Each data
point is obtained by averaging the fraction over 200 different
realizations.

\begin{figure}
\begin{center}
\scalebox{0.4}[0.4]{\includegraphics{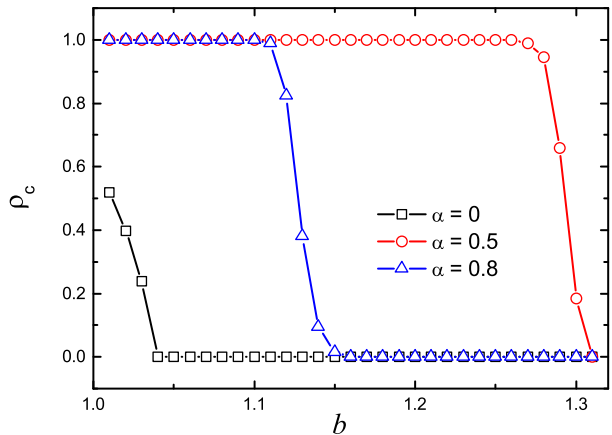}} \caption{(Color
online) Fraction of cooperators $\rho_{c}$ as a function of $b$, the
temptation to defect, for different values of the punishment fine
$\alpha$.} \label{fig1}
\end{center}
\end{figure}

Figure~\ref{fig1} shows the fraction of cooperators $\rho_{c}$ as a
function of $b$, the temptation to defect, for different values of
the punishment fine $\alpha$. We observe, for any given value of
$\alpha$, a monotonic decrease in $\rho_{c}$ as $b$ is increased. In
addition, we find that $\rho_{c}$ can never reach unity in the whole
range of $b$ when the punishment fine is zero. However, for certain
values of $\alpha$, e.g., $\alpha=0.5$ and $\alpha=0.8$, cooperators
can dominate the whole system for $b$ below some critical value.

\begin{figure}
\begin{center}
\scalebox{0.4}[0.4]{\includegraphics{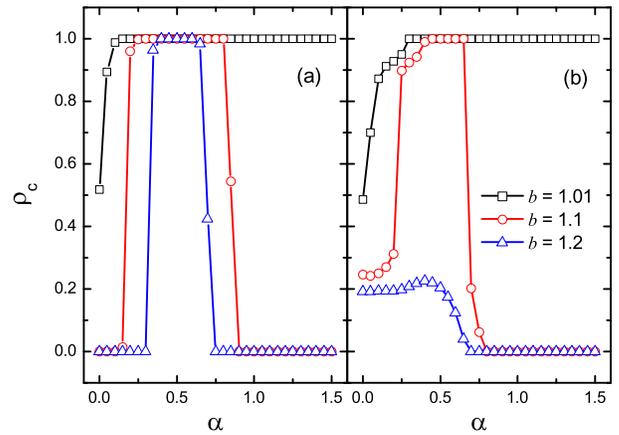}} \caption{(Color
online) Fraction of cooperators $\rho_{c}$ as a function of the
punishment fine $\alpha$ for different values of $b$. The results in
(a) and (b) from simulation and theoretical analysis, respectively.}
\label{fig2}
\end{center}
\end{figure}

\begin{figure}
\begin{center}
\scalebox{0.44}[0.44]{\includegraphics{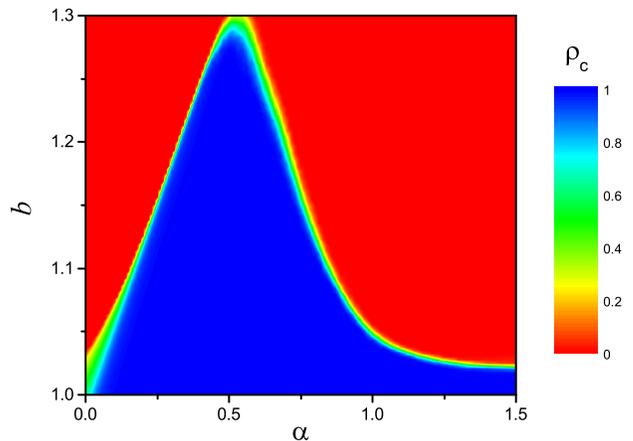}} \caption{(Color
online) Color coded map of the fraction of cooperators $\rho_{c}$ in
the parameter plane ($\alpha$,$b$).} \label{fig3}
\end{center}
\end{figure}

Figure~\ref{fig2} shows $\rho_{c}$ as a function of $\alpha$ for
different values of $b$. We see that, for relatively small values of
$b$ (e.g., $b=1.01$), $\rho_{c}$ increases with $\alpha$. However,
for larger values of $b$ (e.g., $b=1.1$ or $b=1.2$), there exists an
optimal region of $\alpha$ in which full cooperation ($\rho_{c}=1$)
is achieved. For example, the optimal region in $\alpha$ is
approximately $[0.3,0.8]$ and $[0.4,0.6]$ for $b=1.1$ and $b=1.2$
respectively. The optimal value of $\alpha$ is moderate, indicating
that either minor or harsh punishment does not promote cooperation.
The dependence of $\rho_{c}$ on $\alpha$ can be qualitatively
predicted analytically through a pair-approximation
analysis~\cite{update0,ch}, the results from which are shown in
Fig.~\ref{fig2}(b).

To quantify the ability of punishment fine $\alpha$ to promote cooperation
for various values of $b$ more precisely, we compute the behavior of
$\rho_{c}$ in the parameter plane ($\alpha$, $b$), as shown in
Fig.~\ref{fig3}. We see that, for $b<1.02$, $\rho_{c}$ increases to unity
as $\alpha$ is increased. For $1.02 < b < 1.27$, there exists an optimal
region of $\alpha$ in which complete extinction of defectors occurs
($\rho_{c}=1$). The optimal region of $\alpha$ becomes narrow as $b$
is increased. For $b > 1.27$, there also exists an optimal value of
$\alpha$ that results in the highest possible level of cooperation
for the corresponding $b$ values, albeit $\rho_{c}< 1$.

\begin{figure}
\begin{center}
\scalebox{0.4}[0.4]{\includegraphics{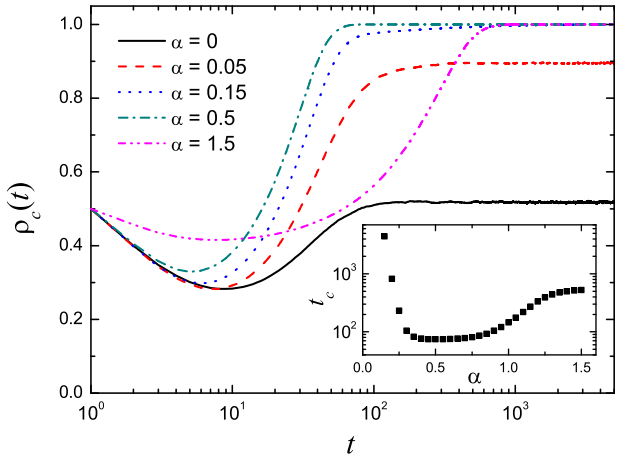}} \caption{(Color
online) For $b = 1.01$, time series of the fraction of cooperators,
$\rho_{c}(t)$, for different values of $\alpha$. The inset presents
the convergence time $t_{c}$ versus $\alpha$.} \label{fig4}
\end{center}
\end{figure}

To gain insights into the mechanism of cooperation enhancement
through punishment, we examine the time evolution of $\rho_{c}$ for
a number of combinations of the parameters $\alpha$ and $b$.
Figure~\ref{fig4} shows the time series $\rho_{c}(t)$ for different
values of $\alpha$ and a relatively small value of $b$ (e.g.,
$b=1.01$). In every case, $\rho_{c}(t)$ decreases initially but then
increases to a constant value. The similar phenomenon was also
observed in Refs.~\cite{a1,a2}. For small values of $\alpha$ (e.g.,
$\alpha=0$ or $\alpha=0.05$), $\rho_{c}(t)$ cannot reach unity. For
relatively large values of $\alpha$ (e.g., $\alpha=0.15$,
$\alpha=0.5$ or $\alpha=1.5$), at the end defectors are extinct and
all individuals are cooperators. We define the convergence time
$t_{c}$ as the number of time steps required for complete extinction
of defectors. In the inset of Fig.~\ref{fig4}, we show $t_{c}$ as a
function of $\alpha$ and observe that $t_{c}$ is minimized for
$\alpha \approx 0.5$.

\begin{figure}
\begin{center}
\scalebox{0.4}[0.4]{\includegraphics{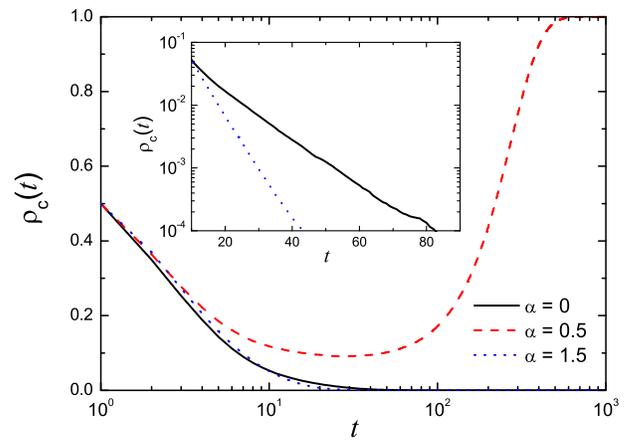}} \caption{(Color
online) For $b = 1.2$, time series $\rho_{c}(t)$ for different
values of $\alpha$. Inset shows that the fraction of cooperators
decays exponentially for $\alpha=0$ and $\alpha=1.5$.} \label{fig5}
\end{center}
\end{figure}

\begin{figure}
\begin{center}
\scalebox{0.42}[0.42]{\includegraphics{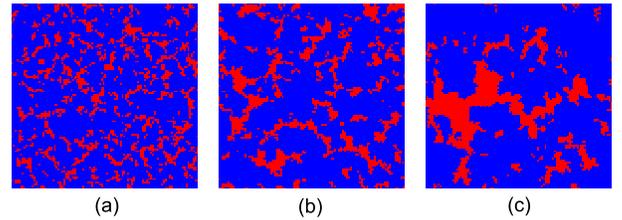}} \caption{(Color
online) For a number of values of $\alpha$, snapshots of typical
distributions of cooperators (blue) and defectors (red) in the
steady state. The fraction of cooperators in the equilibrium state
is set to be $\rho_{c}=0.8$ for different values of $\alpha$. The
values of $\alpha$ and $b$ are (a) $\alpha=0.02$, $b=1.001$; (b)
$\alpha=0.2$, $b=1.116$ and (c) $\alpha=0.4$, $b=1.245$.}
\label{fig6}
\end{center}
\end{figure}

Figure~\ref{fig5} shows the time series $\rho_{c}(t)$ for different
values of $\alpha$ when there is strong temptation to defect (e.g., $b=1.2$).
We observe that cooperators gradually die out for either small
(e.g., $\alpha=0$) or large (e.g., $\alpha=1.5$) $\alpha$ values.
A remarkable phenomenon is that, asymptotically, the fraction of
cooperators decreases exponentially over time for small or large
$\alpha$ values: $\rho_{c}(t)\propto e^{-t/\tau}$, where the value of
$\tau$ depends on $\alpha$, as shown in the inset of Fig.~\ref{fig5}.
For moderate values of $\alpha$ (e.g., $\alpha=0.5$), $\rho_{c}(t)$
decreases initially and then increases to unity.

How the cooperators and defectors are distributed in the physical
space when a steady state is reached? Figure~\ref{fig6} shows
spatial strategy distributions for different values of the
punishment fine $\alpha$ in the equilibrium state. By varying the
value of $b$, we produce the same fraction of cooperators
($\rho_{c}=0.8$) for each value of $\alpha$. We see that, defectors
spread homogeneously in the whole space when $\alpha$ is small
(e.g., $\alpha=0.02$), while the same amount of defectors are more
condensed for the higher value of $\alpha$ (e.g., $\alpha=0.4$).
Such condensation of defectors prevents them to reach competitive
payoffs.

\begin{figure*}
\begin{center}
\scalebox{0.82}[0.82]{\includegraphics{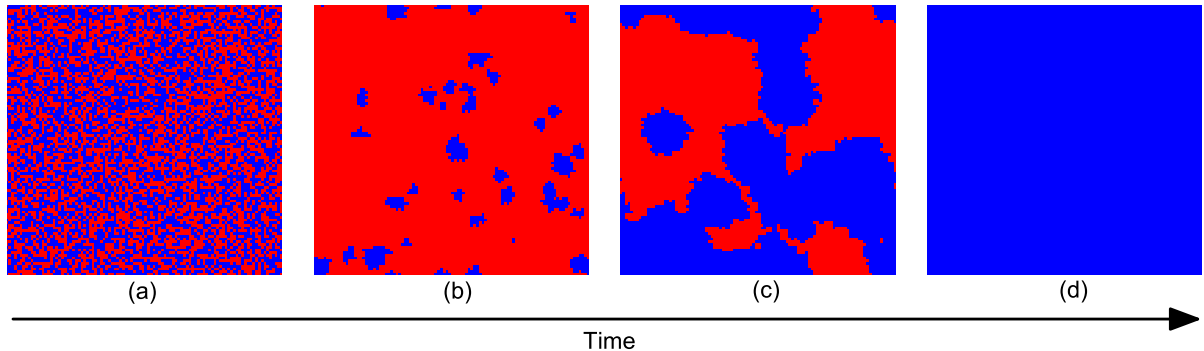}} \caption{(Color
online) For $\alpha = 0.5$ and $b = 1.2$, snapshots of typical
distributions of cooperators (blue) and defectors (red) at different
time steps $t$.} \label{fig7}
\end{center}
\end{figure*}

How does the distribution of cooperators and defectors evolve with
time? Figure~\ref{fig7} shows the distribution of cooperators and
defectors at different time steps for a large value of $b$ (e.g.,
$b=1.2$) and a moderate value of $\alpha$ (e.g., $\alpha=0.5$).
Initially, cooperators and defectors are randomly distributed with
equal probability [Fig.~\ref{fig7}(a)]. After a few time steps,
cooperators and defectors are clustered, and the density of
cooperators is lower than that associated with the initial state
[Fig.~\ref{fig7}(b)]. With time the cooperator clusters continue to
expand and the defector clusters shrink [Fig.~\ref{fig7}(c)].
Finally, the whole population is cooperators [Fig.~\ref{fig7}(d)].
From Fig.~\ref{fig7}, one can also observe that interfaces
separating domains of cooperators and defectors become smooth as
time evolves. As illustrated in Refs.~\cite{Szolnoki,perc}, noisy
borders are beneficial for defectors, while straight domain walls
help cooperators to spread.

\begin{figure}
\begin{center}
\scalebox{0.4}[0.4]{\includegraphics{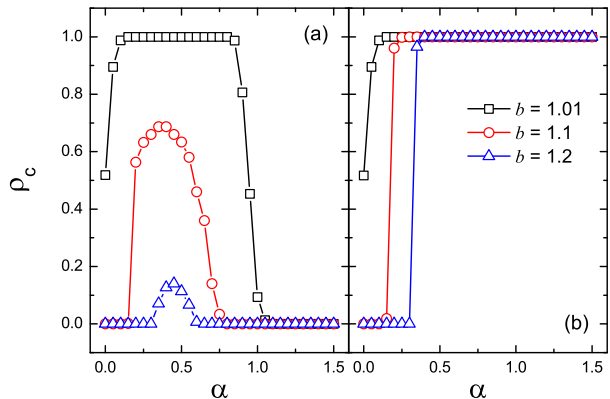}} \caption{(Color
online) Fraction of cooperators $\rho_{c}$ as a function of the
punishment fine $\alpha$ for different values of the temptation to
defect $b$. The initial density of cooperators $\rho_{0}$ is (a) 0.2
and (b) 0.8, respectively.} \label{fig10}
\end{center}
\end{figure}

In the above studies, we set the initial density of cooperators
$\rho_{0}$ to be 0.5. Now we study how different values of
$\rho_{0}$ affect the evolution of cooperation. From
Fig.~\ref{fig10}(a), one can find that for the small value of
$\rho_{0}$ (e.g., $\rho_{0}=0.2$), the cooperation level reaches
maximum at moderate punishment fine when the temptation to defect
$b$ is fixed. However, for the large value of $\rho_{0}$ (e.g.,
$\rho_{0}=0.8$), the cooperation level increases to 1 as the
punishment fine increases [Fig.~\ref{fig10}(b)].

\section{Conclusions and discussions} \label{sec: conclusion}

To obtain quantitative understanding of the role of peer pressure on
cooperation, we study evolutionary game dynamics and propose the
natural mechanism of mutual punishment in which an individual will
punish a neighbor with a fine if their strategies are different, and
vice versa. The mutual punishment can be interpreted as a term
modifying the strength of coordination type
interaction~\cite{szabo}. Because of the symmetry in imposing the
punishment between the individuals, one might expect that it would
have little effect on cooperation. However, we find a number of
counterintuitive phenomena.

In a well-mixed population, if the initial density of cooperators is
no more than 0.5, cooperators die out regardless of the values of
the punishment fine and the temptation to defect. If the initial
density of cooperators exceeds 0.5, for each value of the temptation
to defect, there exists a critical value of the punishment fine,
below (above) which is the full defection (cooperation). The
critical value of the punishment fine increases as the temptation to
defect increases but it decreases as the initial density of
cooperators increases.

For structured population, our main findings are as follows. (i) If
the initial density of cooperators is small (e.g., 0.2), there
exists an optimal value of the punishment fine, leading to the
highest cooperation. Too weak or too harsh punishment will suppress
cooperation. Similar phenomenon was also observed in
Refs.~\cite{the5,jiang}. (ii) If the initial density of cooperators
is moderate (e.g., 0.5), for weak temptation to defect, the final
fraction of cooperators increases to 1 as the punishment fine
increases. For strong temptation to defect, the cooperation level
can be maximized for moderate punishment fine. (iii) If the initial
density of cooperators is large (e.g., 0.8), for each value of the
temptation to defect, the final fraction of cooperators increases to
1 as the punishment fine increases.

In the present studies, we use the prisoner's dilemma game to
understand the role of peer pressure in cooperation. It would be
interesting to explore the effect of mutual punishment on other
types of evolutionary games (e.g., the snowdrift game and the public
goods game) in future work. By our mechanism, an individual can be
punished least by taking the local majority strategy. In fact,
following the majority is an important mechanism for the formation
of public opinion~\cite{majority}. As a side result, our work
provides a connection between the evolutionary games and opinion
dynamics.

\begin{acknowledgments}

This work was supported by the National Natural Science Foundation
of China under Grants No. 61403083, No. 11135001, and No. 11475074.
Y.C.L. was supported by ARO under Grant No. W911NF-14-1-0504.

\end{acknowledgments}

\end{document}